\documentclass[12pt]{article}
\usepackage{latexsym,amssymb,amsmath}
\begin{document}
\baselineskip=20pt

\newcommand{\la}{\langle}
\newcommand{\ra}{\rangle}
\newcommand{\psp}{\vspace{0.4cm}}
\newcommand{\pse}{\vspace{0.2cm}}
\newcommand{\ptl}{\partial}
\newcommand{\dlt}{\delta}
\newcommand{\sgm}{\sigma}
\newcommand{\al}{\alpha}
\newcommand{\be}{\beta}
\newcommand{\G}{\Gamma}
\newcommand{\gm}{\gamma}
\newcommand{\vs}{\varsigma}
\newcommand{\Lmd}{\Lambda}
\newcommand{\lmd}{\lambda}
\newcommand{\td}{\tilde}
\newcommand{\vf}{\varphi}
\newcommand{\yt}{Y^{\nu}}
\newcommand{\wt}{\mbox{wt}\:}
\newcommand{\rd}{\mbox{Res}}
\newcommand{\ad}{\mbox{ad}}
\newcommand{\stl}{\stackrel}
\newcommand{\ol}{\overline}
\newcommand{\ul}{\underline}
\newcommand{\es}{\epsilon}
\newcommand{\dmd}{\diamond}
\newcommand{\clt}{\clubsuit}
\newcommand{\vt}{\vartheta}
\newcommand{\ves}{\varepsilon}
\newcommand{\dg}{\dagger}
\newcommand{\tr}{\mbox{Tr}}
\newcommand{\ga}{{\cal G}({\cal A})}
\newcommand{\hga}{\hat{\cal G}({\cal A})}
\newcommand{\Edo}{\mbox{End}\:}
\newcommand{\for}{\mbox{for}}
\newcommand{\kn}{\mbox{ker}}
\newcommand{\Dlt}{\Delta}
\newcommand{\rad}{\mbox{Rad}}
\newcommand{\rta}{\rightarrow}
\newcommand{\mbb}{\mathbb}
\newcommand{\lra}{\Longrightarrow}

\begin{center}{\Large \bf   Stable-Range Approach to the Equation of}\end{center}
\begin{center}{\Large \bf  Nonstationary Transonic Gas Flows}
\end{center}
\vspace{0.2cm}

\begin{center}{\large Xiaoping Xu}\end{center}
\begin{center}{Institute of Mathematics, Academy of Mathematics \& System Sciences}\end{center}
\begin{center}{Chinese Academy of Sciences, Beijing 100080, P.R. China}
\footnote{Research supported
 by China NSF 10431040}\end{center}

\vspace{0.6cm}

 \begin{center}{\Large\bf Abstract}\end{center}

\vspace{1cm} {\small Using certain finite-dimensional stable range
of the nonlinear terms, we obtain large families of exact
solutions parameterized by functions for the equation of
nonstationary transonic gas flows discovered by Lin, Reisner and
Tsien, and its three-dimensional generalization.} \vspace{0.8cm}

\section{Introduction}

Lin, Reisner and Tsien [LRT] found the  equation
$$2u_{tx}+u_xu_{xx}-u_{yy}=0.\eqno(1.1)$$
for two-dimensional non-steady motion of a slender body in a
compressible fluid, which was later called the ``equation of
nonstationary transonic gas flows" (cf. [M1]). Mamontov  obtained
the Lie point symmetries of the above equation in [M1] and solved
the problem of existence of analytic solutions in [M2]. The
three-dimensional generalization:
$$2u_{tx}+u_xu_{xx}-u_{yy}-u_{zz}=0\eqno(1.2)$$
was studied by Kucharczyk [K] and by Sukhinin [Ss]. Indeed, the
Lie point symmetries of the equation (1.2) were found in their
works. Sevost'janov [Sg] found explicit solutions of the equation
(1.1), describing nonstationary transonic flows in plane nozzles.

In this paper, we present a new approach based on the fact that
the nonlinear terms keep some finite-dimensional polynomial space
in $x$ stable. We obtain a family of solutions of the equation
(1.1) blowing up on a moving line $y=f(t)$,  which reflect partial
phenomena of gust,  and a family of smooth solutions parameterized
by six smooth functions of $t$. Moreover, we find a family of
solutions of the equation (1.2) blowing up on a rotating and
translating plane $\cos\al(t)\:y+\sin\al(t)\:z=f(t)$, which
reflect partial phenomena of turbulence, and a family of solutions
polynomial in $x$ parameterized by time-dependent harmonic
functions in $y$ and $z$, whose special cases are smooth
solutions. In particular, we find all the solutions polynomial in
$x$ and $y$ for the equation (1.1) and all the solutions
polynomial in $x,\;y$ and $z$ for the equation (1.2).  Since our
solutions contain parameter functions, it can be used to solve
certain  boundary-value problems for these equations.

Lie group method is one of most important ways of solving
differential equations. However, the method only enables one to
obtain certain special solutions. It is desirable to find more
effective ways of solving differential equations. Indeed, we do
find one of solving the above nonlinear partial differential
equations.

On Mamonotov's list of the Lie point symmetries of the equation
(1.1) (e.g. cf. Page 296 in [I]), the most sophisticated ones are
those with respect to the following vector fields:
\begin{eqnarray*}\hspace{2cm}& &X_1=3\al(t)\ptl_t+(\al'(t)x+{\al'}'(t)y^2)\ptl_x+2\al'(t)y\ptl_y
\\ &
&+\left[-\al'(t)u+{\al'}'(t)x^2+2{{\al'}'}'(t)xy^2+\frac{1}{3}\al^{(4)}(t)y^4\right]\ptl_u,
\hspace{3cm}(1.3)\end{eqnarray*}
$$X_2=\be'(t)y\ptl_x+\be(t)\ptl_y+\left[2{\be'}'(t)xy+\frac{2}{3}{{\be'}'}'(t)y^3\right]\ptl_u,
\eqno(1.4)$$
$$X_3=\gm(t)\ptl_x+[2\gm'(t)x+2{\gm'}'(t)y^3]\ptl_u,\eqno(1.5)$$
where $\al,\;\be$ and $\gm$ are arbitrary functions of $t$. Among
the known Lie point symmetries of (1.2) (e.g. cf. Page 298 in
[I]), the most interesting ones are those with respect to the
following vector fields:
$$X_4=\frac{5}{2}t^2\ptl_t+\left(tx+\frac{3}{2}(y^2+z^2)\right)\ptl_x+3ty\ptl_y
+3tz\ptl_z+(x^2-3tu)\ptl_u,\eqno(1.6)$$
$$X_5=g'(t)y\ptl_x+g(t)\ptl_y+\left(2{g'}'(t)xy+{{g'}'}'(t)\left(\frac{y^3}{3}+
yz^2\right)\right)\ptl_u,
\eqno(1.7)$$
$$X_6=h'(t)z\ptl_x+h\ptl_z+\left(2{h'}'(t)xz+{{g'}'}'(t)\left(\frac{z^3}{3}+
zy^2\right)\right)\ptl_u,
\eqno(1.8)$$
$$X_7=\al\ptl_x+[2\sgm'(t)x+{\sgm'}'(t)(y^2+z^2)\ptl_u,\eqno(1.9)$$
where $g,\;h,\;\sgm$ and $\al$ are arbitrary functions of $t$.
Ryzhov and Shefter [RS] found some invariant solutions of the
equation (1.2), which represent time-dependent flows in a circular
or plane Laval nozzle.

First we find that the group invariant solutions with respect to
the vector fields $X_1$-$X_7$ are polynomial in $x$ with degree
$\leq 3$. Then we examine the equations (1.1), (1.2) more closely
and observe that this phenomena is essentially caused by the fact
that the nonlinear term $u_xu_{xx}$ keep the following polynomial
subspace stable:
$$\mbb{R}+\mbb{R}x+\mbb{R}x^2+\mbb{R}x^3,\eqno(1.10)$$
where $\mbb{R}$ stands for the field of real numbers. This
observation suggests us a new ansatze of solving the equations
(1.1) and (1.2). Since the equation (1.2) contains the Laplace
operator $\ptl_y^2+\ptl_z^2$, our approach to (1.2) will involve
harmonic analysis and sophisticated integrations. For simplicity,
we will solve the equation (1.1) in Section 1.2 although it can be
viewed as a special case of (1.2). Exact solutions of (1.2) will
be found in Section 3.

\section{Two-Dimensional Case}

In this section, we study solutions polynomial in $x$ for the
Lin-Reisner-Tsien equation (1.1).
 By comparing the terms of highest degree in $x$, we find that
 such a solution must be of the form:
 $$u=f(t,y)+g(t,y)x+h(t,y)x^2+\xi(t,y)x^3,\eqno(2.1)$$
where $f(t,y),\;g(t,y),\;h(t,y)$ and $\xi(t,y)$ are
suitably-differentiable functions to be determined. Note
$$u_x=g+2hx+3\xi x^2,\qquad u_{xx}=2h+6\xi x,\eqno(2.2)$$
$$u_{tx}=g_t+2h_tx+3\xi_t x^2,\qquad
u_{yy}=f_{yy}+g_{yy}x+h_{yy}x^2+\xi_{yy}x^3,\eqno(2.3)$$ Now (1.1)
becomes $$2(g_t+2h_tx+3\xi_t x^2)+(g+2hx+3\xi x^2)(2h+6\xi
x)-f_{yy}-g_{yy}x-h_{yy}x^2-\xi_{yy}x^3=0,\eqno(2.4)$$ which is
equivalent to the following systems of partial differential
equations:
$$\xi_{yy}=18\xi^2,\eqno(2.5)$$
$$h_{yy}=6\xi_t+18\xi h,\eqno(2.6)$$
$$g_{yy}=4h_t+4h^2+6\xi g,\eqno(2.7)$$
$$f_{yy}=2g_t+2gh.\eqno(2.8)$$

First we observe that
$$\xi=\frac{1}{(\sqrt{3}y+\be(t))^2}\eqno(2.9)$$ is a solution of
the equation (2.5) for any differentiable function $\be$ of $t$.
Substituting (2.9) into (2.6), we get
$$h_{yy}=-\frac{12\be'(t)}{(\sqrt{3}y+\be(t))^3}+\frac{18}{(\sqrt{3}y+\be(t))^2}h.\eqno(2.10)$$
Denote by $\mbb{Z}$ the ring of integers. Write
$$h(t,y)=\sum_{i\in\mbb{Z}}a_i(t)(\sqrt{3}y+\be(t))^i.\eqno(2.11)$$
Then
$$h_{yy}=\sum_{i\in\mbb{Z}}3(i+2)(i+1)a_{i+2}(t)(\sqrt{3}y+\be(t))^i.\eqno(2.12)$$
Substituting (2.11) and (2.12) into (2.10), we have
$$\sum_{i\in\mbb{Z}}3[(i+2)(i+1)-6]a_{i+2}(t)(\sqrt{3}y+\be(t))^i=
-\frac{12\be'(t)}{(\sqrt{3}y+\be(t))^3},\eqno(2.13)$$
equivalently,
$$-12a_{-1}(t)=-12\be'(t),\qquad3(i+4)(i-1)a_{i+2}(t)=0,\qquad i\neq -3.\eqno(2.14)$$
Thus
$$h=\frac{\al(t)}{(\sqrt{3}y+\be(t))^2}+\frac{\be'(t)}{\sqrt{3}y+\be(t)}+\gm(t)
(\sqrt{3}y+\be(t))^3,\eqno(2.15)$$ where $\al$ and $\gm$ are
arbitrary  differentiable functions of $t$.

Note
\begin{eqnarray*}h_t&=&-\frac{2\al(t)\be'(t)}{(\sqrt{3}y+\be(t))^3}+
\frac{\al'(t)}{(\sqrt{3}y+\be(t))^2}-\frac{\be'(t)^2}{(\sqrt{3}y+\be(t))^2}\\
&&+\frac{{\be'}'(t)}{\sqrt{3}y+\be(t)}+3\gm(t)\be'(t)
(\sqrt{3}y+\be(t))^2+\gm'(t)
(\sqrt{3}y+\be(t))^3\hspace{2.5cm}(2.16)\end{eqnarray*} and
\begin{eqnarray*}& &h^2=\frac{\al(t)^2}{(\sqrt{3}y+\be(t))^4}+
2\frac{\al(t)\be'(t)}{(\sqrt{3}y+\be(t))^3}+\frac{\be'(t)^2}{(\sqrt{3}y+\be(t))^2}\\
&
&+2\al(t)\gm(t)(\sqrt{3}y+\be(t))+2\gm(t)\be'(t)(\sqrt{3}y+\be(t))^2+
\gm(t)^2(\sqrt{3}y+\be(t))^6.\hspace{1.1cm}(2.17)\end{eqnarray*}
Substituting the above two equations into (2.7), we have:
\begin{eqnarray*}&&g_{yy}-\frac{6}{(\sqrt{3}y+\be)^2}g
 =\frac{4\al^2}{(\sqrt{3}y+\be)^4}+
\frac{4\al'}{(\sqrt{3}y+\be)^2}+\frac{4{\be'}'}{\sqrt{3}y+\be}\\
&&+8\al\gm(\sqrt{3}y+\be)+20\gm\be'(\sqrt{3}y+\be)^2 +4\gm'
(\sqrt{3}y+\be)^3+4\gm^2(\sqrt{3}y+\be)^6.\hspace{1.4cm}(2.18)\end{eqnarray*}
Write
$$g(t,y)=\sum_{i\in\mbb{Z}}b_i(t)(\sqrt{3}y+\be)^i.\eqno(2.19)$$
Then
$$g_{yy}=\sum_{i\in\mbb{Z}}3(i+2)(i+1)b_{i+2}(t)(\sqrt{3}y+\be)^i.\eqno(2.20)$$
Substituting (2.19) and (2.20) into (2.18), we get
$$3(i+3)ib_{i+2}=0,\qquad i\neq -4,-2,-1,1,2,3,6,\eqno(2.21)$$
$$b_{-2}=\frac{\al^2}{3},\;\;b_0=-\frac{2\al'}{3},\;\;b_1=-\frac{2{\be'}'}{3},\;\;
b_3=\frac{2\al\gm}{3},\eqno(2.22)$$
$$b_4=\frac{2\be'\gm}{3},\;\;b_5=\frac{2\gm'}{27},\;\;b_8=\frac{2\gm^2}{81}.\eqno(2.23)$$
Therefore
\begin{eqnarray*}& &g=\frac{\al^2}{3(\sqrt{3}y+\be)^2}+\frac{\sgm}{\sqrt{3}y+\be}-\frac{2\al'}{3}
-\frac{2{\be'}'}{3}(\sqrt{3}y+\be)+\rho(\sqrt{3}y+\be)^2\\ & &
+\frac{2\al\gm}{3}(\sqrt{3}y+\be)^3
+\frac{2\be'\gm}{3}(\sqrt{3}y+\be)^4+\frac{2\gm'}{27}(\sqrt{3}y+\be)^5+\frac{2\gm^2}{81}
(\sqrt{3}y+\be)^8,\hspace{1.1cm}(2.24)\end{eqnarray*} where $\sgm$
and $\rho$ are arbitrary differentiable functions of $t$.

Observe that
\begin{eqnarray*}g_t&=&-\frac{2\al^2\be'}{3(\sqrt{3}y+\be)^3}+
\frac{(2\al\al'-3\sgm\be')}{3(\sqrt{3}y+\be)^2}+\frac{\sgm'}{\sqrt{3}y+\be}-\frac{2{\al'}'}{3}
-\frac{2\be'{\be'}'}{3}\\ &
&+\frac{2(3\rho\be'-{{\be'}'}')}{3}(\sqrt{3}y+\be)+(\rho'+2\al\gm\be')(\sqrt{3}y+\be)^2\\
& & + \frac{2\al'\gm+2\al\gm'+8(\be')^2\gm}{3}(\sqrt{3}y+\be)^3
+\frac{18{\be'}'\gm+28\be'\gm'}{27}(\sqrt{3}y+\be)^4\\ &
&+\frac{2{\gm'}'}{27}(\sqrt{3}y+\be)^5 +\frac{16\gm^2\be'}{81}
(\sqrt{3}y+\be)^7 +\frac{4\gm\gm'}{81}
(\sqrt{3}y+\be)^8,\hspace{3.2cm}(2.25)\end{eqnarray*}
\begin{eqnarray*}&
&gh=\frac{\al^3}{3(\sqrt{3}y+\be)^4}+\frac{3\al\sgm+\al^2\be'}{3(\sqrt{3}y+\be)^3}
+\frac{3\be'\sgm-2\al\al'}{3(\sqrt{3}y+\be)^2}-\frac{2(\al{\be'}'+\al'\be')}{3(\sqrt{3}y+\be)}
\\ &
&+\al\rho-\frac{2\be'{\be'}'}{3}+(\al^2\gm+\be'\rho)(\sqrt{3}y+\be)
+\frac{4\al\be'\gm+3\gm\sgm}{3}(\sqrt{3}y+\be)^2\\ &
&+\frac{18(\be')^2\gm-16\al\gm'}{27}(\sqrt{3}y+\be)^3+\frac{2\be'\gm'-18{\be'}'\gm}{27}
(\sqrt{3}y+\be)^4+\gm\rho(\sqrt{3}y+\be)^5\\
&&+\frac{56\al\gm^2}{81} (\sqrt{3}y+\be)^6+\frac{56\be'\gm^2}{81}
(\sqrt{3}y+\be)^7+\frac{2\gm^3}{81}
(\sqrt{3}y+\be)^{11}.\hspace{3.5cm}(2.26)\end{eqnarray*}
Substituting (2.25) and (2.26) into (2.8), we obtain
\begin{eqnarray*}&
&f_{yy}=\frac{2\al^3}{3(\sqrt{3}y+\be)^4}
+\frac{6\al\sgm-2\al^2\be'}{3(\sqrt{3}y+\be)^3}
+\frac{6\sgm'-4(\al{\be'}'+\al'\be')}{3(\sqrt{3}y+\be)}
\\ &&+2\al\rho-\frac{4{\al'}'}{3}-\frac{8\be'{\be'}'}{3}
+\frac{6\al^2\gm+18\be'\rho-4{{\be'}'}'}{3}(\sqrt{3}y+\be)\\ & &
+\frac{20\al\be'\gm+6\gm\sgm+6\rho'}{3}(\sqrt{3}y+\be)^2
+\frac{180(\be')^2\gm+4\al\gm'+36\al'\gm}{27}(\sqrt{3}y+\be)^3
\hspace{6cm}\end{eqnarray*}\begin{eqnarray*}&
&+\frac{20\be'\gm'}{9}
(\sqrt{3}y+\be)^4+\frac{54\gm\rho+4{\gm'}'}{27}(\sqrt{3}y+\be)^5
+\frac{112\al\gm^2}{81}
(\sqrt{3}y+\be)^6\\&&+\frac{16\be'\gm^2}{9} (\sqrt{3}y+\be)^7+
\frac{8\gm\gm'}{81} (\sqrt{3}y+\be)^8 +\frac{4\gm^3}{81}
(\sqrt{3}y+\be)^{11}.\hspace{3.5cm}(2.27)\end{eqnarray*} Thus
\begin{eqnarray*}&
&f=\frac{\al^3}{27(\sqrt{3}y+\be)^2}
+\frac{3\al\sgm-\al^2\be'}{9(\sqrt{3}y+\be)}+\theta+\vartheta
y+\al\rho y^2-\frac{2{\al'}'+4\be'{\be'}'}{3}y^2
\\&
&+\frac{1}{9}[6\sgm'-4(\al{\be'}'+\al'\be')](\sqrt{3}y+\be)[\ln(\sqrt{3}y+\be)-1]
+\frac{3\al^2\gm+9\be'\rho-2{{\be'}'}'}{27}(\sqrt{3}y+\be)^3
\\ & &+\frac{10\al\be'\gm+3\gm\sgm+3\rho'}{54}(\sqrt{3}y+\be)^4
+\frac{45(\be')^2\gm+\al\gm'+9\al'\gm}{405}(\sqrt{3}y+\be)^5
\\ & &+\frac{2\be'\gm'}{81}
(\sqrt{3}y+\be)^6+\frac{27\gm\rho+2{\gm'}'}{1701}(\sqrt{3}y+\be)^7
+\frac{2\al\gm^2}{243} (\sqrt{3}y+\be)^8\\ & &
+\frac{2\be'\gm^2}{243} (\sqrt{3}y+\be)^9+\frac{4\gm\gm'}{10935}
(\sqrt{3}y+\be)^{10} +\frac{\gm^3}{9477}
(\sqrt{3}y+\be)^{13},\hspace{3.2cm}(2.28)\end{eqnarray*} where
$\theta$ and $\vartheta$ are arbitrary functions of $t$. \psp

{\bf Theorem 2.1}. {\it We have the following solution of the
equation (1.1) blowing up on the surface $\sqrt{3}y+\be(t)=0$:
\begin{eqnarray*}&
&u=\frac{x^3}{(\sqrt{3}y+\be)^2}+\frac{\al x^2}{(\sqrt{3}y+\be)^2}
+\frac{\be'x^2}{\sqrt{3}y+\be}+\gm (\sqrt{3}y+\be)^3x^2\\
&
&+[\frac{\al^2}{3(\sqrt{3}y+\be)^2}+\frac{\sgm}{\sqrt{3}y+\be}-\frac{2\al'}{3}
-\frac{2{\be'}'}{3}(\sqrt{3}y+\be)+\rho(\sqrt{3}y+\be)^2\\ & &
+\frac{2\al\gm}{3}(\sqrt{3}y+\be)^3
+\frac{2\be'\gm}{3}(\sqrt{3}y+\be)^4+\frac{2\gm'}{27}(\sqrt{3}y+\be)^5+\frac{2\gm^2}{81}
(\sqrt{3}y+\be)^8]x\\ & & +\frac{\al^3}{27(\sqrt{3}y+\be)^2}
+\frac{3\al\sgm-\al^2\be'}{9(\sqrt{3}y+\be)}+\theta+\vartheta
y+\al\rho y^2-\frac{2{\al'}'+4\be'{\be'}'}{3}y^2
\\&
&+\frac{1}{9}[6\sgm'-4(\al{\be'}'+\al'\be')](\sqrt{3}y+\be)[\ln(\sqrt{3}y+\be)-1]
+\frac{3\al^2\gm+9\be'\rho-2{{\be'}'}'}{27}(\sqrt{3}y+\be)^3
\\ & &+\frac{10\al\be'\gm+3\gm\sgm+3\rho'}{54}(\sqrt{3}y+\be)^4
+\frac{45(\be')^2\gm+\al\gm'+9\al'\gm}{405}(\sqrt{3}y+\be)^5
\\ & &+\frac{2\be'\gm'}{81}
(\sqrt{3}y+\be)^6+\frac{27\gm\rho+2{\gm'}'}{1701}(\sqrt{3}y+\be)^7
+\frac{2\al\gm^2}{243} (\sqrt{3}y+\be)^8\\ & &
+\frac{2\be'\gm^2}{243} (\sqrt{3}y+\be)^9+\frac{4\gm\gm'}{10935}
(\sqrt{3}y+\be)^{10} +\frac{\gm^3}{9477}
(\sqrt{3}y+\be)^{13},\hspace{3.3cm}(2.29)\end{eqnarray*} where
$\al,\be,\gm,\sgm,\rho,\theta$ and $\vartheta$ are arbitrary
functions of $t$, whose derivatives appeared in the above exist in
a certain open set of $\mbb{R}$.}\psp

When $\al=\gm=\sgm=\rho=\theta=\vartheta=0$, the above solution
becomes $$u=\frac{x^3}{(\sqrt{3}y+\be)^2}
+\frac{\be'x^2}{\sqrt{3}y+\be} -\frac{2{\be'}'}{3}(\sqrt{3}y+\be)x
-\frac{4\be'{\be'}'}{3}y^2
-\frac{2{{\be'}'}'}{27}(\sqrt{3}y+\be)^3.\eqno(2.30)$$ \pse

Take the trivial solution $\xi=0$ of (2.5), which is the only
solution polynomial in $y$. Then (2.6) and (2.7) become
$$h_{yy}=0,\qquad g_{yy}=4h_t+4h^2.\eqno(2.31)$$
Thus
$$h=\al(t)+\be(t)y.\eqno(2.32)$$
Hence
$$g_{yy}=4\al^2+4\al'+4(\be'+2\al\be)y+4\be^2 y^2.\eqno(2.33)$$
So
$$g=\gm+\sgm y+2(\al^2+\al')y^2+\frac{2}{3}(\be'+2\al\be)y^3+\frac{1}{3}\be^2
y^4,\eqno(2.34)$$ where $\gm$ and $\sgm$ are arbitrary functions
of $t$. Now (2.8) yields
\begin{eqnarray*} f_{yy}&=&2(\al\gm+\gm')+2(\al\sgm+\be\gm+\sgm')y+2(\be\sgm+2\al^3+6\al\al'+2{\al'}')y^2
\\
&&+\frac{20\al^2\be+12\al\be'+20\al'\be+4{\be'}'}{3}y^3+\frac{10\al\be^2+8\be\be'}{3}y^4+
\frac{2}{3}\be^3 y^5.\hspace{2.2cm}(2.35)\end{eqnarray*}
Therefore,
\begin{eqnarray*} f&=&\tau+\rho y+(\al\gm+\gm')y^2+\frac{1}{3}(\al\sgm+\be\gm+\sgm')y^3
+\frac{1}{6}(\be\sgm+2\al^3+6\al\al'+2{\al'}')y^4
\\
&&+\frac{5\al^2\be+3\al\be'+5\al'\be+{\be'}'}{15}y^5+\frac{5\al\be^2+4\be\be'}{45}y^6+
\frac{1}{63}\be^3 y^7.\hspace{3.3cm}(2.36)\end{eqnarray*} \pse

{\bf Theorem 2.2}. {\it The following is a solution of the
equation (1.1):
\begin{eqnarray*} u&=&(\al+\be y)x^2+\left[\gm+\sgm y+2(\al^2+\al')y^2+\frac{2}{3}(\be'+2\al\be)y^3+\frac{1}{3}\be^2
y^4\right]x\\ & &+ \tau+\rho
y+(\al\gm+\gm')y^2+\frac{1}{3}(\al\sgm+\be\gm+\sgm')y^3
+\frac{1}{6}(\be\sgm+2\al^3+6\al\al'+2{\al'}')y^4
\\
&&+\frac{5\al^2\be+3\al\be'+5\al'\be+{\be'}'}{15}y^5+\frac{5\al\be^2+4\b4\be'}{45}y^6+
\frac{1}{63}\be^3 y^7,\hspace{3.2cm}(2.37)\end{eqnarray*} where
$\al,\be,\gm,\sgm,\rho$ and $\tau$ are arbitrary functions of $t$,
whose derivatives appeared in the above exist in a certain open
set of $\mbb{R}$. Moreover, any solution polynomial in $x$ and $y$
of (1.1) must be of the above form. The above solution is smooth
(analytic) if all $\al,\be,\gm,\sgm,\rho$ and $\tau$ are smooth
(analytic) functions of $t$. }\psp

{\bf Remark 2.3}. In addition to the nonzero solution (2.9) of the
equation (2.5), the other nonzero solutions  are of the form
$$\xi=\wp_\iota(\sqrt{3}y+\be(t)),\eqno(2.38)$$
where $\wp_\iota(w)$ is the Weierstrass's elliptic function such
that
$$\wp'_\iota(w)^2=4(\wp_\iota(w)^3-\iota),\eqno(2.39)$$
and $\iota$ is a nonzero constant and $\be$ is any function of
$t$. When $\be$ is not a constant, the solutions of (2.6)-(2.8)
are extremely complicated. If $\be$ is constant, we can take
$\be=0$ by adjusting $\iota$. Any solution of (2.6)-(2.8) with
$h\neq 0$ is also very complicated. Thus the only simple solution
of the equation (1.1) in this case is
$$u=\wp_\iota(\sqrt{3}y)\:x^3.\eqno(2.40)$$

\section{Three-Dimensional Case}

By comparing the terms of highest degree, we find that
 a  solution polynomial in $x$ of the equation (1.2) must be of the form:
$$u=f(t,y,z)+g(t,y,z)x+h(t,y,z)x^2+\xi(t,y,z)x^3,\eqno(3.1)$$
 where $f(t,y,z),\;g(t,y,z),\;h(t,y,z)$ and $\xi(t,y,z)$ are
suitably-differentiable functions to be determined. Note
$$u_x=g+2hx+3\xi x^2,\qquad u_{xx}=2h+6\xi x,\eqno(3.2)$$
$$u_{tx}=g_t+2h_tx+3\xi_t x^2,\qquad
u_{yy}=f_{yy}+g_{yy}x+h_{yy}x^2+\xi_{yy}x^3,\eqno(3.3)$$
$$u_{zz}=f_{zz}+g_{zz}x+h_{zz}x^2+\xi_{zz}x^3.\eqno(3.4)$$
Now (1.2) becomes
\begin{eqnarray*}\hspace{1cm}& &2(g_t+2h_tx+3\xi_t x^2)+(g+2hx+3\xi x^2)(2h+6\xi
x)-(f_{yy}+f_{zz})\\ &
&-(g_{yy}+g_{zz})x-(h_{yy}+h_{zz})x^2-(\xi_{yy}+\xi_{zz})x^3=0,\hspace{4.4cm}(3.5)\end{eqnarray*}
which is equivalent to the following systems of partial
differential equations:
$$\xi_{yy}+\xi_{zz}=18\xi^2,\eqno(3.6)$$
$$h_{yy}+h_{zz}=6\xi_t+18\xi h,\eqno(3.7)$$
$$g_{yy}+g_{zz}=4h_t+4h^2+6\xi g,\eqno(3.8)$$
$$f_{yy}+f_{zz}=2hg_t+2gh.\eqno(3.9)$$

First we observe that
$$\xi=\frac{1}{(\sqrt{3}(y\cos \al(t)+z\sin
\al(t))+\be(t))^2}\eqno(3.10)$$ is a solution of the equation
(3.6), where $\al$ and $\be$ are suitable differentiable functions
of $t$. With the above $\xi$, (3.7) becomes
$$h_{yy}+h_{zz}=-\frac{12(\sqrt{3}\al'(-y\sin
\al+z\cos\al)+\be')}{(\sqrt{3}(y\cos \al+z\sin
\al)+\be)^3}+\frac{18h}{(\sqrt{3}(y\cos \al+z\sin
\al)+\be)^2}.\eqno(3.11)$$ In order to solve (3.11), we change
variables:
$$\zeta=\sqrt{3}(\cos \al\:y+\sin
\al\:z)+\be,\;\;\eta=\sqrt{3}(-\sin
\al\:y+\cos\al\:z).\eqno(3.12)$$ Then
$$\ptl_y=\sqrt{3}(\cos \al\:\ptl_\zeta-\sin
\al\:\ptl_\eta),\qquad\ptl_z=\sqrt{3}(\sin
\al\:\ptl_\zeta+\cos\al\:\ptl_\eta).\eqno(3.13)$$ Thus
$$\ptl_y^2+\ptl_z^2=3(\cos \al\:\ptl_\zeta-\sin
\al\:\ptl_\eta)^2+3(\sin
\al\:\ptl_\zeta+\cos\al\:\ptl_\eta)^2=3(\ptl_\zeta^2+\ptl_\eta^2).\eqno(3.14)$$
Note
$$\ptl_t(\zeta)=\al'\eta+\be',\qquad\ptl_t(\eta)=\al'(\be-\zeta).\eqno(3.15)$$

The equation (3.11) can be rewritten as:
$$h_{\zeta\zeta}+h_{\eta\eta}=-4(\al'\eta+\be')\zeta^{-3}+6\zeta^{-2}h.\eqno(3.16)$$
In order to solve the above equation, we assume
$$h=\sum_{i\in\mbb{Z}}a_i(t,\eta)\zeta^i.\eqno(3.17)$$
Now (3.16) becomes
$$\sum_{i\in\mbb{Z}}a_{i\eta\eta}\zeta^i+
\sum_{i\in\mbb{Z}}i(i-1)a_i\zeta^{i-2}=-4(\al'\eta+\be')\zeta^{-3}
+6\sum_{i\in\mbb{Z}}a_i\zeta^{i-2},\eqno(3.18)$$ which is
equivalent to
$$a_{-3\eta\eta}+2a_{-1}=-4(\al'\eta+\be')+6a_{-1},\;\;
a_{i\eta\eta}+(i+2)(i+1)a_{i+2}=6a_{i+2}\eqno(3.19)$$ for $-3\neq
i\in\mbb{Z}$. Hence
$$a_{-1}=\frac{1}{4}a_{-3\eta\eta}+\al'\eta+\be',\;\;(i+4)(i-1)a_{i+2}=-a_{i\eta\eta}
\qquad\for\;\;-3\neq i\in\mbb{Z}.\eqno(3.20)$$ When $i=-4$ and
$i=1$, we get $a_{-4\eta\eta}=a_{1\eta\eta}=0$. Moreover, $a_{-2}$
and $a_3$ can be any functions.

 Take
$$a_3=\sgm,\;\;
a_{-2}=\rho,\;\;a_{-1}=\al'\eta+\be',\eqno(3.21)$$
$$a_1=a_{-1-2i}=a_{-2-2i}=0\qquad\for\;\;0<i\in\mbb{Z}\eqno(3.22)$$
in order to avoid infinite number of negative powers of $\zeta$ in
(3.17), where $\sgm$ and $\rho$ are are functions of $t$ and
$\eta$ differentiable in a certain domain. By (3.20),
$$a_{3+2k}=\frac{(-1)^k\ptl_{\eta}^{2k}(\sgm)}{\prod_{i=1}^k(2i+5)(2i)}
=\frac{(-1)^k15\ptl_{\eta}^{2k}(\sgm)}
{(2k+5)(2k+3)(2k+1)!},\eqno(3.23)$$
$$a_{-2+2k}=\frac{(-1)^k\ptl_{\eta}^{2k}(\rho)}{\prod_{i=1}^k(2i)(2i-5)}=
\frac{(-1)^k(2k-1)(2k-3)\ptl_{\eta}^{2k}(\rho)}
{3(2k)!}.\eqno(3.24)$$ Therefore,
\begin{eqnarray*}\hspace{1.5cm}h&=&(\al'\eta+\be')\zeta^{-1}+\sum_{k=0}^\infty(-1)^k
[\frac{15\ptl_{\eta}^{2k}(\sgm)\zeta^3} {(2k+5)(2k+3)(2k+1)!}\\ &
& +\frac{(2k-1)(2k-3)\ptl_{\eta}^{2k}(\rho)\zeta^{-2}}
{3(2k)!}]\zeta^{2k}\hspace{6.4cm}(3.25)\end{eqnarray*} is a
solution of (3.16).

By (3.12) and (3.14), (3.8) is equivalent to
$$g_{\zeta\zeta}+g_{\eta\eta}=\frac{4}{3}h_t+\frac{4}{3}h^2+2\zeta^{-2} g.\eqno(3.26)$$
Note
\begin{eqnarray*}& &h_t=({\al'}'\eta+{\be'}'+(\al')^2\be)\zeta^{-1}-(\al')^2-(\al'\eta+\be')^2
\zeta^{-2} +\sum_{k=0}^\infty(-1)^k\zeta^{2k}\{\\ & &\times
\left(\frac{15\ptl_{\eta}^{2k}(\sgm_t+\al'(\be-\zeta)\sgm_\eta)\zeta^3}
{(2k+5)(2k+3)(2k+1)!}+\frac{(2k-1)(2k-3)\ptl_{\eta}^{2k}(\rho_t+\al'(\be-\zeta)\rho_\eta)\zeta^{-2}}
{3(2k)!}\right)\\
&&+(\al'\eta+\be') [\frac{15\ptl_{\eta}^{2k}(\sgm)\zeta^2}
{(2k+5)(2k+1)!}+\frac{(2k-1)(2k-2)(2k-3)\ptl_{\eta}^{2k}
(\rho)\zeta^{-3}}{3(2k)!}]\}.\hspace{1.7cm}(3.27)\end{eqnarray*}
For convenience of solving the equation (3.26), we denote
$$\frac{4}{3}h_t+\frac{4}{3}h^2=\sum_{i=-4}^\infty
b_i(t,\eta)\zeta^i\eqno(3.28)$$ by (3.25) and (3.27). In
particular,
$$b_{-4}=\frac{4}{3}\rho^2,\qquad b_{-3}=0,\eqno(3.29)$$
$$b_{-2}=\frac{4}{3}(\rho_t+\al'\be\rho_\eta)+\frac{4}{9}\rho_{\eta\eta}\rho,\eqno(3.30)$$
$$b_{-1}=\frac{4}{3}[{\al'}'\eta+{\be'}'+(\al')^2\be-\al'\rho_\eta]
+\frac{4}{9}(\al'\eta+\be')\rho_{\eta\eta},\eqno(3.31)$$
$$b_0=-\frac{4(\al')^2}{3}
+\frac{2}{9}(\rho_{t\eta\eta}+\al'\beta\rho_{\eta\eta\eta})+\frac{1}{9}
\ptl^4_\eta(\rho)\rho+\frac{1}{27}\rho^2_{\eta\eta}.\eqno(3.32)$$

Suppose that
$$g=\sum_{i\in\mbb{Z}}c_i(t,\eta)\zeta^i\eqno(3.33)$$
is a solution (3.26). Then
$$\sum_{i\in\mbb{Z}}[i(i-1)c_i\zeta^{i-2}+c_{i\eta\eta}\zeta^i]=\sum_{r=-4}^\infty
b_r\zeta^r+\sum_{i\in\mbb{Z}}2c_i\zeta^{i-2},\eqno(3.34)$$
equivalently
$$(i+3)ic_{i+2}=b_i-c_{i\eta\eta},\;\;(r+3)rc_{r+2}=-c_{r\eta\eta},\qquad
r<-4\leq i.\eqno(3.35)$$ By the second equation in (3.35), we take
$$c_r=0\qquad\for\;\;r<-4\eqno(3.36)$$
to avoid infinite number of negative powers of $\zeta$ in (3.33).
Letting $i=-3,0$, we get
$$b_{-3}=c_{-3\eta\eta},\qquad b_0=c_{0\eta\eta}.\eqno(3.37)$$

The first equation is naturally satisfied because
$c_{-3}=-c_{-5\eta\eta}/10=0$. Taking $i=-2,-4$ and $r=-6$ in
(3.35), we obtain
$$c_0=\frac{1}{2}c_{-2\eta\eta}-\frac{1}{2}b_{-2},\qquad c_{-2}=\frac{1}{4}b_{-4}.
\eqno(3.38)$$ So
$$c_0=\frac{1}{8}\ptl_{\eta}^2(b_{-4})-\frac{1}{2}b_{-2}.\eqno(3.39)$$
Thus we get a constraint:
$$b_0=\frac{1}{8}\ptl_{\eta}^4(b_{-4})-\frac{1}{2}\ptl_\eta^2(b_{-2}),\eqno(3.40)$$
equivalently,
\begin{eqnarray*} \hspace{2cm}& &-\frac{4(\al')^2}{3}
+\frac{2}{9}(\rho_{t\eta\eta}+\al'\beta\rho_{\eta\eta\eta})+\frac{1}{9}
\ptl^4_\eta(\rho)\rho+\frac{1}{27}\rho^2_{\eta\eta}\\
&=&\frac{1}{6}\ptl_{\eta}^4(\rho^2)-
\frac{2}{3}(\rho_{t\eta\eta}+\al'\be\rho_{\eta\eta\eta})-\frac{2}{9}\ptl_\eta^2
(\rho_{\eta\eta}\rho).\hspace{4.5cm}(3.41)\end{eqnarray*} Thus
$$48(\rho_{t\eta\eta}+\al'\beta\rho_{\eta\eta\eta})+6
\ptl^4_\eta(\rho)\rho+
2\rho^2_{\eta\eta}-9\ptl_{\eta}^4(\rho^2)+12\ptl_\eta^2
(\rho_{\eta\eta}\rho)=72(\al')^2.\eqno(3.42)$$

It can be proved by considering the terms of highest degree that
any solution  of (3.42) polynomial in $\eta$ must be of the form
 $$\rho=\gm_0(t)+\gm_1(t)\eta+\gm_2(t)\eta^2.\eqno(3.43)$$
Then (3.42) becomes
$$12\gm_2'-20\gm_2^2=9(\al')^2.\eqno(3.44)$$
So
$$\al'=\frac{2\es}{3}\sqrt{3\gm_2'-5\gm_2^2}\Rightarrow\al=\frac{2\es}{3}\int
\sqrt{3\gm_2'-5\gm_2^2}dt,\eqno(3.45)$$ where $\es=\pm 1$. Replace
$\be$ by $-\be$ if necessary, we can take $\es=1$.
 Under the assumption
(3.43),
$$h=\rho\zeta^{-2}+(\al'\eta+\be')\zeta^{-1}+\frac{\gm_2}{6}+\sum_{k=0}^\infty(-1)^k
\frac{15\ptl_{\eta}^{2k}(\sgm)\zeta^{3+2k}}{(2k+5)(2k+3)(2k+1)!}\eqno(3.46)$$
and
$$b_{-2}=\frac{4}{3}(\rho_t+\al'\be\rho_\eta)+\frac{8}{9}\gm_2\rho,\eqno(3.47)$$
$$b_{-1}=\frac{4}{3}[{\al'}'\eta+{\be'}'+(\al')^2\be-\al'\rho_\eta]
+\frac{8}{9}(\al'\eta+\be')\gm_2,\eqno(3.48)$$
$$b_0=-\frac{4(\al')^2}{3}
+\frac{4}{9}\gm_2'+\frac{4}{27}\gm_2^2\eqno(3.49)$$ by
(3.30)-(3.32).

 Denote
$$\Psi_{\la\be,\rho,\sgm\ra}(t,\eta,\zeta)=\sum_{i=1}^\infty
b_i\zeta^i.\eqno(3.50)$$ For any real function $F(t,\eta)$
analytic at $\eta=\eta_0$, we define
$$F(t,\eta_0+\sqrt{-1}\zeta)=\sum_{r=0}^\infty\frac{\ptl^r_\eta(F)(t,\eta_0)}{r!}
(\sqrt{-1}\zeta)^r. \eqno(3.51)$$ Note
\begin{eqnarray*}& &\sum_{k=0}^\infty(-1)^k
\frac{15\ptl_{\eta}^{2k}(\sgm)\zeta^{3+2k}}{(2k+5)(2k+3)(2k+1)!}
=15\zeta^2\int_0^\zeta\left(\sum_{k=0}^\infty(-1)^k
\frac{\ptl_{\eta}^{2k}(\sgm)\tau_1^{2k}}{(2k+5)(2k+3)(2k)!}\right)d\tau_1\\
&=&15\int_0^\zeta\tau_2\int_0^{\tau_2}\left(\sum_{k=0}^\infty(-1)^k
\frac{\ptl_{\eta}^{2k}(\sgm)\tau_1^{2k}}{(2k+5)(2k)!}\right)d\tau_1\:d\tau_2\\
&=&15\zeta^{-2}\int_0^\zeta\tau_3\int_0^{\tau_3}\tau_2\int_0^{\tau_2}\left(\sum_{k=0}^\infty(-1)^k
\frac{\ptl_{\eta}^{2k}(\sgm)\tau_1^{2k}}{(2k)!}\right)d\tau_1\:d\tau_2\:d\tau_3\\
&=&\frac{15}{2}\zeta^{-2}\int_0^\zeta\tau_3\int_0^{\tau_3}\tau_2\int_0^{\tau_2}[\sgm(t,\eta+
\sqrt{-1}\tau_1)+\sgm(t,\eta-\sqrt{-1}\tau_1)]d\tau_1\:d\tau_2\:d\tau_3
,\hspace{1.3cm}(3.52)\end{eqnarray*}
\begin{eqnarray*}& &\ptl_t
\left[\sum_{k=0}^\infty(-1)^k
\frac{15\ptl_{\eta}^{2k}(\sgm)\zeta^{3+2k}}{(2k+5)(2k+3)(2k+1)!}\right]
\\ &=&
\sum_{k=0}^\infty(-1)^k
\frac{15\ptl_{\eta}^{2k}(\sgm_t)\zeta^{3+2k}}{(2k+5)(2k+3)(2k+1)!}+\al'\be\sum_{k=0}^\infty(-1)^k
\frac{15\ptl_{\eta}^{2k+1}(\sgm)\zeta^{3+2k}}{(2k+5)(2k+3)(2k+1)!}
\\ & &-\al'\sum_{k=0}^\infty(-1)^k
\frac{15\ptl_{\eta}^{2k+1}(\sgm)\zeta^{4+2k}}{(2k+5)(2k+3)(2k+1)!}+(\al'\eta+\be')
\sum_{k=0}^\infty(-1)^k
\frac{15\ptl_{\eta}^{2k}(\sgm)\zeta^{2+2k}}{(2k+5)(2k+1)!}
\hspace{6cm}\end{eqnarray*}\begin{eqnarray*} &=&
\frac{15}{2}\zeta^{-2}\int_0^\zeta\tau_3\int_0^{\tau_3}\tau_2\int_0^{\tau_2}[\sgm_t(t,\eta+
\sqrt{-1}\tau_1)+\sgm_t(t,\eta-\sqrt{-1}\tau_1)]d\tau_1\:d\tau_2\:d\tau_3
\\ &&+\frac{15\al'(\zeta-\be)}{2\zeta^2}\sqrt{-1}\int_0^\zeta\tau_2\int_0^{\tau_2}\tau_1[\sgm(t,\eta+
\sqrt{-1}\tau_1)-\sgm(t,\eta-\sqrt{-1}\tau_1)] d\tau_1\:d\tau_2\\
&&+\frac{15}{2}(\al'\eta+\be')
\zeta^{-3}\int_0^\zeta\tau_2^3\int_0^{\tau_2}[\sgm(t,\eta+
\sqrt{-1}\tau_1)+\sgm(t,\eta-\sqrt{-1}\tau_1)] d\tau_1\:d\tau_2.
\hspace{1cm}(3.53)\end{eqnarray*} According to (3.28) and (3.50),
we have
\begin{eqnarray*}& &\Psi_{\la\be,\rho,\sgm\ra}(t,\eta,\zeta)=\\ & &
75\zeta^{-4}\left(\int_0^\zeta\tau_3\int_0^{\tau_3}\tau_2\int_0^{\tau_2}[\sgm(t,\eta+
\sqrt{-1}\tau_1)+\sgm(t,\eta-\sqrt{-1}\tau_1)]d\tau_1\:d\tau_2\:d\tau_3\right)^2\\&
&
+10\zeta^{-2}\int_0^\zeta\tau_3\int_0^{\tau_3}\tau_2\int_0^{\tau_2}[\sgm_t(t,\eta+
\sqrt{-1}\tau_1)+\sgm_t(t,\eta-\sqrt{-1}\tau_1)]d\tau_1\:d\tau_2\:d\tau_3
\\ &&+\frac{10\al'(\zeta-\be)}{\zeta^2}\sqrt{-1}
\int_0^\zeta\tau_2\int_0^{\tau_2}\tau_1[\sgm(t,\eta+
\sqrt{-1}\tau_1)-\sgm(t,\eta-\sqrt{-1}\tau_1)] d\tau_1\:d\tau_2\\
&&+10(\al'\eta+\be')
\zeta^{-3}\int_0^\zeta\tau_2^3\int_0^{\tau_2}[\sgm(t,\eta+
\sqrt{-1}\tau_1)+\sgm(t,\eta-\sqrt{-1}\tau_1)] d\tau_1\:d\tau_2\\
& &
+20\left(\int_0^\zeta\tau_3\int_0^{\tau_3}\tau_2\int_0^{\tau_2}[\sgm(t,\eta+
\sqrt{-1}\tau_1)+\sgm(t,\eta-\sqrt{-1}\tau_1)]d\tau_1\:d\tau_2\:d\tau_3\right)\\&&\times
\zeta^{-2}\left(\rho\zeta^{-2}+(\al'\eta+\be')\zeta^{-1}+\frac{\gm_2}{6}\right).
\hspace{7.8cm}(3.54)\end{eqnarray*}

Now
$$c_{-2}=\frac{\rho^2}{3}\eqno(3.55)$$
by (3.29) and (3.38). According to (3.35) with $i=-3,0$, $c_{-1}$
and $c_2$ can be arbitrary. For convenience, we redenote
$$ c_{-1}=\kappa(t,\eta),\qquad c_2=\omega(t,\eta).\eqno(3.56)$$
 Moreover, (3.29),  (3.39) and (3.47) imply
$$c_0=\frac{\rho_\eta^2}{3}-
\frac{2}{3}(\rho_t+\al'\be\rho_\eta)+\frac{2}{9}\gm_2\rho.\eqno(3.57)$$
Furthermore, (3.31) and (3.35),
$$c_1=\frac{\kappa_{\eta\eta}}{2}-\frac{2}{3}[{\al'}'\eta+{\be'}'+(\al')^2\be-\al'\rho_\eta]
-\frac{4}{9}(\al'\eta+\be')\gm_2.\eqno(3.58)$$ In addition, (3.35)
and (3.56) yield $$c_{2k+3}=
\frac{(-1)^{k+1}\ptl_\eta^{2k+4}(\kappa)}{2(k+2)(2k+2)!}+
\sum_{i=0}^k\frac{(-1)^{k-i}(i+1)(2i)!}{(k+2)(2k+2)!}\ptl_\eta^{2(k-i)}(b_{2i+1}),
\eqno(3.59)$$
$$c_{2k+4}=\frac{(-1)^{k+1}3\ptl_\eta^{2k+2}(\omega)}{(2k+5)(2k+3)!}
+\sum_{i=0}^k\frac{(-1)^{k-i}(2i+3)(2i+1)!}{(2k+5)(2k+3)!}\ptl_\eta^{2(k-i)}(b_{2i+2})
\eqno(3.60)$$ for $0\leq k\in\mbb{Z}$.

Set
\begin{eqnarray*}& &\Phi_{\la\be,\rho,\sgm,\kappa,\omega\ra}(t,\eta,\zeta)=
\kappa\zeta^{-1}+
\frac{\kappa_{\eta\eta}\zeta}{2}+\omega\zeta^2+\sum_{i=3}^\infty
c_i\zeta^i\\
&=&-\zeta\ptl_\zeta\zeta^{-1}
\left[\sum_{k=0}^\infty(-1)^k\frac{\ptl_\eta^{2k}(\kappa)\zeta^{2k}}{(2k)!}\right]
+\zeta^2\sum_{k=0}^\infty(-1)^k\frac{3\ptl_\eta^{2k}(\omega)\zeta^{2k}}{(2k+3)(2k+1)!}
\\& &+\sum_{k=0}^\infty
\sum_{i=0}^k\frac{(-1)^{k-i}(i+1)(2i)!}{(k+2)(2k+2)!}\ptl_\eta^{2(k-i)}(b_{2i+1})\zeta^{2k+3}
\\ & &+\sum_{k=0}^\infty\sum_{i=0}^k\frac{(-1)^{k-i}(2i+3)(2i+1)!}{(2k+5)(2k+3)!}
\ptl_\eta^{2(k-i)}(b_{2i+2})\zeta^{2k+4}.\hspace{4.3cm}(3.61)\end{eqnarray*}
Note
\begin{eqnarray*}\hspace{1cm} & &\zeta^2\sum_{k=0}^\infty(-1)^k\frac{3\ptl_\eta^{2k}(\omega)\zeta^{2k}}{(2k+3)(2k+1)!}
\\ &=&\frac{3}{2}\zeta^{-1}\int_0^\zeta\tau_2\int_0^{\tau_2}[\omega(t,\eta+\sqrt{-1}\tau_1)
+\omega(t,\eta-\sqrt{-1}\tau_1)]d\tau_1\:d\tau_2.\hspace{2.2cm}(3.62)\end{eqnarray*}
Moreover,
$$\Psi_{\la\be,\rho,\sgm\ra}(t,\eta,0)=0,\qquad b_i=
\frac{\ptl_\zeta^i(\Psi_{\la\be,\rho,\sgm\ra})(t,\eta,0)}{i!}\qquad\for\;\;0<i\in\mbb{Z}.
\eqno(3.63)$$ Thus
\begin{eqnarray*}& &\Phi_{\la\be,\rho,\sgm,\kappa,\omega\ra}(t,\eta,\zeta)=
\frac{3}{2}\zeta^{-1}\int_0^\zeta\tau_2\int_0^{\tau_2}[\omega(t,\eta+\sqrt{-1}\tau_1)
+\omega(t,\eta-\sqrt{-1}\tau_1)]d\tau_1\:d\tau_2\\ & &
-\frac{1}{2}\zeta\ptl_\zeta\zeta^{-1}[\kappa(t,\eta+\sqrt{-1}\tau_1)
+\kappa(t,\eta-\sqrt{-1}\tau_1)]
\\& &+\sum_{k=0}^\infty
\sum_{i=0}^k\frac{(-1)^{k-i}(i+1)\ptl_\eta^{2(k-i)}\ptl_\zeta^{2i+1}(\Psi_{\la\be,\rho,\sgm\ra})(t,\eta,0)}
{(2i+1)(k+2)(2k+2)!}\zeta^{2k+3}
\\ & &+\sum_{k=0}^\infty\sum_{i=0}^k\frac{(-1)^{k-i}(2i+3)
\ptl_\eta^{2(k-i)}\ptl_\zeta^{2i+2}(\Psi_{\la\be,\rho,\sgm\ra})(t,\eta,0)
}{(2i+2)(2k+5)(2k+3)!}
\zeta^{2k+4},\hspace{3.5cm}(3.64)\end{eqnarray*} in which the
summations are finite if $\sgm(t,\eta)$ is polynomial in $\eta$.
Now
\begin{eqnarray*}& &h=
\rho\zeta^{-2}+(\al'\eta+\be')\zeta^{-1}+\frac{\gm_2}{6}
+\frac{15}{2}\zeta^{-2}\\ & &\times
\int_0^\zeta\tau_3\int_0^{\tau_3}\tau_2\int_0^{\tau_2}[\sgm(t,\eta+
\sqrt{-1}\tau_1)+\sgm(t,\eta-\sqrt{-1}\tau_1)]d\tau_1\:d\tau_2\:d\tau_3,
\hspace{2.2cm}(3.65)\end{eqnarray*}
\begin{eqnarray*}g&=&\Phi_{\la\be,\rho,\sgm,\kappa,\omega\ra}(t,\eta,\zeta)+
\frac{\rho^2}{3}\zeta^{-2} +\frac{\rho_\eta^2}{3}-
\frac{2}{3}(\rho_t+\al'\be\rho_\eta)+\frac{2}{9}\gm_2\rho\\ & &
-\frac{2}{3}[{\al'}'\eta+{\be'}'+(\al')^2\be-\al'\rho_\eta]\zeta
-\frac{4}{9}(\al'\eta+\be')\gm_2\zeta.\hspace{4.8cm}(3.66)\end{eqnarray*}

Denote
$$\Lmd_{\la\be,\rho,\sgm,\kappa,\omega\ra}(t,\eta,\zeta)=
\frac{2}{3}(gh+\ptl_t(g)).\eqno(3.67)$$ Then the equation (3.9)
becomes
$$f_{\eta\eta}+f_{\zeta\zeta}=\Lmd_{\la\be,\rho,\sgm,\kappa,\omega\ra}
(t,\eta,\zeta).\eqno(3.68)$$ Set
$$w=\frac{\eta+\sqrt{-1}\zeta}{2},\qquad\ol{w}=\frac{\eta-\sqrt{-1}\zeta}{2}.
\eqno(3.69)$$ Then
$$\ptl_\eta=\frac{1}{2}(\ptl_w+\ptl_{\ol{w}}),\qquad
\ptl_\zeta=\frac{\sqrt{-1}}{2}(\ptl_w-\ptl_{\ol{w}}).\eqno(3.70)$$
Hence
$$\ptl_\eta^2+\ptl_\zeta^2=\ptl_w\ptl_{\ol{w}}.\eqno(3.71)$$

 A complex function
$$G(\mu)\;\;\mbox{is
called {\it bar-homomorphic}
if}\;\;\ol{G(\mu)}=G(\ol{\mu}).\eqno(3.72)$$ For instance,
trigonometric functions, polynomials with real coefficients and
elliptic functions with bar-invariant periods are bar-homomorphic
functions. The extended function $F(t,\mu)$ in (3.51) is
bar-homomorphic in $\mu$. Now we have
\begin{eqnarray*}f&=&\int_{w_0}^w\int_{\ol{w_0}}^{\ol{w}}
\Lmd_{\la\be,\rho,\sgm,\kappa,\omega\ra}(t,\mu+\ol{\mu},\sqrt{-1}(\ol{\mu}-\mu))d\ol{\mu}\:d\mu
+\chi(t,\eta+\sqrt{-1}\zeta)
\\ & &+\chi(t,\eta-\sqrt{-1}\zeta)+\sqrt{-1}[\nu(t,\eta+\sqrt{-1}\zeta)
-\nu(t,\eta-\sqrt{-1}\zeta)],\hspace{2.9cm}(3.73)\end{eqnarray*}
where $w_0$ is a fixed complex number and
$\chi(t,\mu),\;\nu(t,\mu)$ are complex functions in real variable
$t$ and bar-homomorphic in complex variable $\mu$. \psp

{\bf Theorem 3.1}. {\it In terms of  the notions in (3.12), the
function $u=\zeta^{-2}x^3+hx^2+gx+f$ is a solution of the equation
(1.2) blowing up on the hypersurface $\sqrt{3}(\cos \al\:y+\sin
\al\:z)+\be=0$, with $h$ given in (3.65), $g$ given in (3.66) via
(3.54) and (3.64), and $f$ given in (3.73) via (3.67). The
involved parametric functions $\rho$ is given in (3.43), $\al$ is
given in (3.45), and $\sgm,\;\kappa,\;\omega$ are real functions
in real variable $t$ and $\eta$.} \psp

Next we want to find a more explicit formula when
$\sgm=\kappa=\omega=\chi=\nu=0$. In this case,
$$h=\rho\zeta^{-2}+(\al'\eta+\be')\zeta^{-1}+\frac{\gm_2}{6}\eqno(3.74)$$
and
\begin{eqnarray*}\hspace{2cm}g&=&\frac{\rho^2}{3}\zeta^{-2}
+\frac{\rho_\eta^2}{3}-
\frac{2}{3}(\rho_t+\al'\be\rho_\eta)+\frac{2}{9}\gm_2\rho\\ & &
-\frac{2}{3}[{\al'}'\eta+{\be'}'+(\al')^2\be-\al'\rho_\eta]\zeta
-\frac{4}{9}(\al'\eta+\be')\gm_2\zeta.\hspace{2.9cm}(3.75)\end{eqnarray*}
 Moreover, (3.15) and (3.43) yield
\begin{eqnarray*}& &g_t+hg=\frac{\rho^3}{3}\zeta^{-4}-\frac{(\al'\eta+\be')\rho^2}{3}\zeta^{-3}
+\frac{5\gm_2\rho^2+6\rho\rho_\eta^2}{18}\zeta^{-2}
+[\frac{\al'\eta+\be'}{9}(3\rho_\eta^2\\ &
&-6\rho_t-6\al'\be\rho_\eta-2\gm_2\rho)
-\frac{2}{3}\rho({\al'}'\eta+{\be'}'+(\al')^2\be)]\zeta^{-1}
+\frac{2\rho_\eta\rho_{t\eta}}{3}+\frac{2\gm_2'\rho}{9}\\
&
&-\frac{2}{3}(\rho_{tt}+2\al'\be\rho_{t\eta}+(\al'\be)'\rho_\eta+2(\al'\be)^2\gm_2)
+\frac{\gm_2\rho_t}{9}+ \frac{13\al'\be\gm_2\rho_\eta}{9}
 +\frac{\gm_2\rho_\eta^2}{18}+\frac{\gm_2^2\rho}{27}\\ & &
-\frac{4}{3}(\al'\eta+\be')({\al'}'\eta+{\be'}'+(\al')^2\be-\al'\rho_\eta)
-\frac{8}{9}\gm_2(\al'\eta+\be')^2+[\frac{2\al'\rho_{t\eta}}{3}\\
& &+
\frac{\gm_2}{9}(19(\al')^2\be-5{\al'}'\eta-5{\be'}')-\frac{13\al'\gm_2\rho_\eta}{9}
-\frac{2}{27} \gm_2^2(\al'\eta+\be') -\frac{2}{3}(
{{\al'}'}'\eta+{{\be'}'}'\\ &
&+(\al')^2\be'-{\al'}'\rho_\eta+3\al'{\al'}'\be)
-\frac{4}{9}(\al'\gm_2'\eta+\be'\gm_2')]\zeta
+[\frac{2\al'{\al'}'}{3}-\frac{8(\al')^2\gm_2}{9}]\zeta^2.\hspace{1.4cm}(3.76)\end{eqnarray*}
By (3.67) and (3.68), we can find $f$.\psp

{\bf Corollary 3.2}. {\it  In terms of  the notions in (3.12), we
have the following solution of the equation (1.2) which blows up
on the hypersurface $\sqrt{3}(\cos \al\:y+\sin \al\:z)+\be=0$:
\begin{eqnarray*} &
&u=\zeta^{-2}x^3+\left[\rho\zeta^{-2}+(\al'\eta+\be')\zeta^{-1}+\frac{\gm_2}{6}\right]x^2+[
\frac{\rho^2}{3}\zeta^{-2} +\frac{\rho_\eta^2}{3}-
\frac{2}{3}(\rho_t+\al'\be\rho_\eta)\\ & &+\frac{2}{9}\gm_2\rho
-\frac{2}{3}[{\al'}'\eta+{\be'}'+(\al')^2\be-\al'\rho_\eta]\zeta
-\frac{4}{9}(\al'\eta+\be')\gm_2\zeta]x\\ & &
+\frac{\rho^3}{27}\zeta^{-2}
-\frac{(\al'\eta+\be')\rho^2}{9}\zeta^{-1}+\frac{\gm_2\rho^2}{27}\ln\zeta
-\frac{\gm_2\rho^2_\eta+2\gm_2^2\rho}{54}\zeta^2(2\ln\zeta-3)
\\ & &+\frac{\gm_2^3}{324}\zeta^4(12\ln\zeta-25)
+[\frac{4}{9}\al'\rho\rho_\eta+\frac{2(\al'\eta+\be')}{27}(9\rho_\eta^2-6\rho_t
-6\al'\be\rho_\eta+4\gm_2\rho)
\\ & &-\frac{4}{9}\rho({\al'}'\eta+{\be'}'+(\al')^2\be)]
\zeta(\ln\zeta-1)
-\frac{2}{243}[29\al'\gm_2\rho_\eta+(\al'\eta+\be')(20\gm_2^2-3\gm_2')
\\ &
&-9(\al')^2\be\gm_2-3(\al'\rho_{t\eta}+\gm_2({\al'}'\eta+{\be'}')+{\al'}'\rho_\eta)]
\zeta^3(6\ln\zeta-11)+[\frac{2\rho_\eta\rho_{t\eta}}{9}+\frac{2\gm_2'\rho}{27}\\
&
&-\frac{2}{9}(\rho_{tt}+2\al'\be\rho_{t\eta}+(\al'\be)'\rho_\eta+2(\al'\be)^2\gm_2)
+\frac{\gm_2\rho_t}{27}+ \frac{13\al'\be\gm_2\rho_\eta}{27}
 +\frac{\gm_2\rho_\eta^2}{54}\\ & &+\frac{\gm_2^2\rho}{81}
-\frac{4}{9}(\al'\eta+\be')({\al'}'\eta+{\be'}'+(\al')^2\be-\al'\rho_\eta)
-\frac{8}{27}\gm_2(\al'\eta+\be')^2]\zeta^2
\\ & &+[\frac{2\al'\rho_{t\eta}}{27}+
\frac{\gm_2}{81}(19(\al')^2\be-5{\al'}'\eta-5{\be'}')-\frac{13\al'\gm_2\rho_\eta}{81}
-\frac{2}{243} \gm_2^2(\al'\eta+\be') \\ & &-\frac{2}{27}(
{{\al'}'}'\eta+{{\be'}'}'+(\al')^2\be'-{\al'}'\rho_\eta+3\al'{\al'}'\be)
-\frac{4}{81}(\al'\gm_2'\eta+\be'\gm_2')]\zeta^3
 \\ & &-[
\frac{\gm_2\gm_2'}{6}-\frac{{\gm_2'}'}{27} +\frac{7\gm_2^3}{486}-
\frac{\al'{\al'}'}{9}
+\frac{4(\al')^2\gm_2}{27}]\zeta^4,\hspace{6.4cm}(3.77)\end{eqnarray*}
where $\rho$ is a third-order differentiable function  as in
(3.43), $\al$ is given in (3.45) and $\be$ is any third-order
differentiable function of $t$.}\psp

Finally,  $\xi=0$ is the only solution of the equation (3.6)
polynomial in $y$ and $z$. Under this assumption, the equations
(3.7) and (3.8) becomes
$$h_{yy}+h_{zz}=0,\qquad g_{yy}+g_{zz}=4h_t+4h^2.\eqno(3.78)$$
The first equation is a Laplace equation whose solutions are
called {\it harmonic functions}. It can be proved as (3.25) by
power series that the general solution of the first equation is:
$$h=(\sgm+\sqrt{-1}\rho)(t,y+\sqrt{-1}z)+(\sgm-\sqrt{-1}\rho)(t,y-\sqrt{-1}z),\eqno(3.79)$$
where $\sgm(t,\mu)$ and $\rho(t,\mu)$ are complex functions in
real variable $t$ and bar-homomorphic in complex variable $\mu$
(cf. (3.72)). Set
$$w=y+\sqrt{-1}z,\qquad \ol{w}=y-\sqrt{-1}z.\eqno(3.80)$$
Then the Laplace operator
$$\ptl_y^2+\ptl_z^2=4\ptl_w\ptl_{\ol{w}}.\eqno(3.81)$$
The second equation in (3.78) is equivalent to:
\begin{eqnarray*}\hspace{1cm}&&\ptl_w\ptl_{\ol{w}}(g)=h_t+h^2=(\sgm_t+\sqrt{-1}\rho_t)(t,w)
+(\sgm_t-\sqrt{-1}\rho_t)(t,\ol{w})\\ &
&+[(\sgm+\sqrt{-1}\rho)(t,w)+(\sgm-\sqrt{-1}\rho)(t,\ol{w})]^2.
\hspace{5.5cm}(2.82)\end{eqnarray*} Hence the general solution of
the second equation in (3.78) is:
\begin{eqnarray*}&&g=\int_{\ol{w_1}}^{\ol{w}}\int_{w_1}^{w}\{(\sgm_t+\sqrt{-1}\rho_t)(t,\mu_1)
+(\sgm_t-\sqrt{-1}\rho_t)(t,\ol{\mu_1})\\ &
&+[(\sgm+\sqrt{-1}\rho)(t,\mu_1)+(\sgm-\sqrt{-1}\rho)(t,\ol{\mu_1})]^2\}d\mu_1\:d\ol{\mu_1}\\
& &+(\kappa+\sqrt{-1}\omega)(t,w)
+(\kappa-\sqrt{-1}\omega)(t,\ol{w}),\hspace{6.8cm}(3.83)\end{eqnarray*}
where $\kappa(t,\mu)$ and $\omega(t,\mu)$ are complex functions in
real variable $t$ and bar-homomorphic in complex variable $\mu$
(cf. (3.72)), and $w_1$ is a complex constant. Furthermore, (3.9)
becomes
$$\ptl_w\ptl_{\ol{w}}(f)=\frac{1}{2}(g_t+gh).\eqno(3.84)$$

Observe
\begin{eqnarray*}& &g_t=\int_{\ol{w_1}}^{\ol{w}}\int_{w_1}^{w}[
2((\sgm+\sqrt{-1}\rho)(t,\mu_1)+(\sgm-\sqrt{-1}\rho)(t,\ol{\mu_1}))
((\sgm_t+\sqrt{-1}\rho_t)(t,\mu_1)\\ & &
+(\sgm_t-\sqrt{-1}\rho_t)(t,\ol{\mu_1}))+(\sgm_{tt}+\sqrt{-1}\rho_{tt})(t,\mu_1)+
(\sgm_{tt}-\sqrt{-1}\rho_{tt})(t,\ol{\mu_1})]d\mu_1\:d\ol{\mu_1}
\\ & &+(\kappa_t+\sqrt{-1}\omega_t)(t,w)
+(\kappa_t-\sqrt{-1}\omega_t)(t,\ol{w}).\hspace{6.5cm}(3.85)\end{eqnarray*}
Thus the general solution of the equation (3.9) is:
\begin{eqnarray*}&&f=\frac{1}{2}\int_{\ol{w_2}}^{\ol{w}}\int_{w_2}^{w}\{
\int_{\ol{w_1}}^{\ol{\mu_2}}\int_{w_1}^{\mu_2}[
2((\sgm+\sqrt{-1}\rho)(t,\mu_1)+(\sgm-\sqrt{-1}\rho)(t,\ol{\mu_1}))\\
& &\times
((\sgm_t+\sqrt{-1}\rho_t)(t,\mu_1)+(\sgm_t-\sqrt{-1}\rho_t)(t,\ol{\mu_1}))+
(\sgm_{tt}+\sqrt{-1}\rho_{tt})(t,\mu_1)\\
& &+
(\sgm_{tt}-\sqrt{-1}\rho_{tt})(t,\ol{\mu_1})]d\mu_1\:d\ol{\mu_1}
+(\kappa_t+\sqrt{-1}\omega_t)(t,\mu_2)
+(\kappa_t-\sqrt{-1}\omega_t)(t,\ol{\mu_2})\\ &
&+[(\sgm+\sqrt{-1}\rho)(t,\mu_2)+(\sgm-\sqrt{-1}\rho)(t,\ol{\mu_2})]
[(\kappa+\sqrt{-1}\omega)(t,\mu_2)
\\ & &+(\kappa-\sqrt{-1}\omega)(t,\ol{\mu_2})+
\int_{\ol{w_1}}^{\ol{\mu_2}}\int_{w_1}^{\mu_2}\{(\sgm_t+\sqrt{-1}\rho_t)(t,\mu_1)
+(\sgm_t-\sqrt{-1}\rho_t)(t,\ol{\mu_1})\\ &
&+[(\sgm+\sqrt{-1}\rho)(t,\mu_1)+(\sgm-\sqrt{-1}\rho)(t,\ol{\mu_1})]^2\}
d\mu_1\:d\ol{\mu_1}] \} d\mu_2\:d_{\ol{\mu_2}}\\
&&+(\chi+\sqrt{-1}\nu)(t,w)
+(\chi-\sqrt{-1}\nu)(t,\ol{w}),\hspace{6.9cm}(3.86)\end{eqnarray*}
where $\chi(t,\mu)$ and $\nu(t,\mu)$ are complex functions in real
variable $t$ and bar-homomorphic in complex variable $\mu$, and
$w_2$ is a complex constant. \psp

{\bf Theorem 3.3}. {\it In terms of the notions in (3.79), the
following is a solution polynomial in $x$ of the equation (1.2):
\begin{eqnarray*}&
&u=[(\sgm+\sqrt{-1}\rho)(t,w)+(\sgm-\sqrt{-1}\rho)(t,\ol{w})]x^2+\{
\int_{\ol{w_1}}^{\ol{w}}\int_{w_1}^{w}\{(\sgm_t+\sqrt{-1}\rho)(t,\mu_1)
\\ & &+(\sgm_t-\sqrt{-1}\rho)(t,\ol{\mu_1})+[(\sgm+\sqrt{-1}\rho)
(t,\mu_1)+(\sgm-\sqrt{-1}\rho)(t,\ol{\mu_1})]^2\}d\mu_1\:d\ol{\mu_1}\\
& & +(\kappa+\sqrt{-1}\omega)(t,w)
+(\kappa-\sqrt{-1}\omega)(t,\ol{w})\}x+
\frac{1}{2}\int_{\ol{w_2}}^{\ol{w}}\int_{w_2}^{w}\{
\int_{\ol{w_1}}^{\ol{\mu_2}}\int_{w_1}^{\mu_2}\\ & &[
2((\sgm+\sqrt{-1}\rho)(t,\mu_1)
+(\sgm-\sqrt{-1}\rho)(t,\ol{\mu_1}))
((\sgm_t+\sqrt{-1}\rho_t)(t,\mu_1)\\
&
&+(\sgm_t-\sqrt{-1}\rho_t)(t,\ol{\mu_1}))+(\sgm_{tt}+\sqrt{-1}\rho_{tt})(t,\mu_1)+
(\sgm_{tt}-\sqrt{-1}\rho_{tt})(t,\ol{\mu_1})]d\mu_1\:d\ol{\mu_1}
\\ &&+(\kappa_t+\sqrt{-1}\omega_t)(t,\mu_2)
+(\kappa_t-\sqrt{-1}\omega_t)(t,\ol{\mu_2})+[(\sgm+\sqrt{-1}\rho)(t,\mu_2)\\
& &+(\sgm-\sqrt{-1}\rho)(t,\ol{\mu_2})]
[(\kappa+\sqrt{-1}\omega)(t,\mu_2)
+(\kappa-\sqrt{-1}\omega)(t,\ol{\mu_2})\\ & &+
\int_{\ol{w_1}}^{\ol{\mu_2}}\int_{w_1}^{\mu_2}\{(\sgm_t+\sqrt{-1}\rho_t)(t,\mu_1)
+(\sgm_t-\sqrt{-1}\rho_t)(t,\ol{\mu_1})\\ &
&+[(\sgm+\sqrt{-1}\rho)(t,\mu_1)+(\sgm-\sqrt{-1}\rho)(t,\ol{\mu_1})]^2\}
d\mu_1\:d\ol{\mu_1}]\} d\mu_2\:d_{\ol{\mu_2}}\\
&&+(\chi+\sqrt{-1}\nu)(t,w)
+(\chi-\sqrt{-1}\nu)(t,\ol{w}),\hspace{6.9cm}(3.87)\end{eqnarray*}
 where
$\sgm(t,\mu),\;\rho(t,\mu),\;\kappa(t,\mu),\;\omega(t,\mu),\;\chi(t,\mu)$
and $\nu(t,\mu)$  complex functions in real variable $t$ and
bar-homomorphic in complex variable $\mu$ (cf. (3.72)). Moreover,
the above solution is smooth (analytic) if all
$\sgm,\;\rho,\;\kappa,\;\omega,\;\chi$ and $\nu$ are smooth
(analytic) functions. In particular, any solution of the equation
(1.2) polynomial in $x,y,z$ must be of the form (3.87) in which
$\sgm,\;\rho,\;\kappa,\;\omega,\;\chi$ and $\nu$ are polynomial in
$\mu$.}\psp

{\bf Remark 3.4}. In addition to the solutions in Theorems 3.1 and
3.2, the equation (1.2) has the following simple solution:
$$u=\wp_\iota(\sqrt{3}(ay+bz))\:x^3,\eqno(3.88)$$
where $\wp_\iota(w)$ is the Weierstrass's elliptic function
satisfying (2.39) and $a,b$ are real constants such that
$a^2+b^2=1$.

\vspace{0.7cm}

\noindent{\Large \bf References}

\begin{description}

\item[{[I]}] N. H. Ibragimov, {\it Lie Group Analysis of
Differential Equations}, Volume 1, CRC Handbook, CRC Press, 1995.

\item[{[LRT]}] C. C. Lin, E. Reissner and H. S. Tsien, On
two-dimensional non-steady motion of a slender body in a
compressible fluid, {\it J. Math, Phys.} {\bf 27} (1948), no. 3,
220

\item[{[Kp1]}] P. Kucharczyk, Group properties of the ``short
waves" equations in gas dynamics, {\it Bull. Acad. Polon. Sci.,
Ser. Sci. Techn.} {\bf XIII} (1965), no. 5, 469

\item[{[Kp2]}] P. Kucharczyk, {\it Teoria Grup Liego w
Zastosowaniu do R\'{o}wman R\'{o}zniczkowych Czaskowych,} IPPT
Polish Academy of Sciences, Warsaw, 65, 1967.

\item[{[M1]}] E. V. Mamontov, On the theory of nonstationary
transonic flows, {\it Dokl. Acad. Nauk SSSR} {\bf 185} (1969), no.
3, 538

\item[{[M2]}] E. V. Mamontov, Analytic perturbations in a
nonstationary transonic stream, {\it Dinamika Splosn. Sredy Vyp.}
{\bf 10} (1972), 217-222.

\item[{[RS]}] O. S. Ryzhov and G. M. Shefter, On unsteady gas
flows in Laval nozzles, {\it Soviet Physics. Dokl.} {\bf 4}
(1959), 939-942.

\item[{[Sg]}] G. D. Sevost'janov, An equation for nonstationary
transonic flows of an ideal gas, {\it Izv. Acad. Nauk SSSR Meh.
Zidk. Gaza}, 1977, no.1, 105-109.

\item[{[Sv]}] S. V. Sukhinin, Group property and conservation laws
of the equation of transonic motion of gas, {\it Dinamika Splosh.
Sredi} {\bf 36} (1978), 130.

\end{description}

\end{document}